\documentclass[runningheads]{llncs}
\usepackage[T1]{fontenc}
\usepackage{graphicx}
\usepackage{tablefootnote}
\usepackage{multirow}
\usepackage{subcaption}
\usepackage{booktabs}
\usepackage{array}
\usepackage{adjustbox}
\usepackage{caption}
\usepackage{comment}
\usepackage{placeins}
\usepackage{xcolor}
\newcommand{\joann}{\textcolor{magenta}}
\newcommand{\gw}{\textcolor{blue}}
\newcommand{\sigmasq}{\ensuremath{\sigma^2}}

\begin{document}
\title{A Study on the Data Distribution Gap \\ in Music Emotion Recognition}
%
%
\author{Joann Ching\inst{1}\orcidID{0009-0000-8355-1067} \and
Gerhard Widmer\inst{1,2}\orcidID{0000-0003-3531-1282}}
\authorrunning{J. Ching et al.}
%
\institute{Institute of Computational Perception, Johannes Kepler University Linz, Austria \\
\and
LIT AI Lab, Linz Institute of Technology, Linz, Austria \\
\email{\{first.last\}@jku.at}}
\maketitle              
\begin{abstract}
Music Emotion Recognition (MER) is a task deeply connected to human perception, relying heavily on subjective annotations collected from contributors. Prior studies tend to focus on specific musical styles rather than incorporating a diverse range of genres, such as rock and classical, within a single framework. In this paper, we address the task of recognizing emotion from audio content by investigating five datasets with dimensional emotion annotations -- EmoMusic, DEAM, PMEmo, WTC, and WCMED -- which span various musical styles. 
We demonstrate the problem of out-of-distribution generalization in a systematic experiment.
By closely looking at multiple data and feature sets, we provide insight into genre-emotion relationships in existing data and examine potential genre dominance and dataset biases in certain feature representations. Based on these experiments, we arrive at a simple yet effective framework that combines embeddings extracted from the Jukebox model with chroma features and demonstrate how, alongside a combination of several diverse training sets, this permits us to train models with substantially improved cross-dataset generalization capabilities.

\end{abstract}

\section{Introduction}\label{sec:introduction}
The relationship between music and emotion has been a long-investigated topic in MIR. Music emotions can be categorized into two types: ``perceived'' and ``induced'' \cite{Yang_Chen_2011}. \textit{Perceived emotion} refers to the emotion conveyed by the music itself \cite{gabrielsson2002perceive}, while \textit{induced emotion} describes the feelings invoked in the listener \cite{kreutz2007induce}. In this work, we focus solely on perceived emotion.

Music Emotion Recognition (MER) research generally follows two approaches: (1) regression-based \cite{yang2008regression,cheuk2020triplet,malheiro2018lyricemo,koh2021deepaudio} or (2)  classification-based \cite{Yang2006MusicEC,song2012evalEmo,koh2021deepaudio}. These approaches correspond to different annotation formats, which are either \textit{dimensional} (e.g., valence-arousal) or \textit{categorical (or discrete)} (e.g., happy or sad) \cite{ekman1999basic}. Russell's \cite{russell1980circumplex} circumplex model, with its two proposed dimensions, valence (positivity of emotional responses) and arousal (emotional intensity), is often considered less ambiguous than categorical labels \cite{yang2008regression}, as it captures emotional variation on a continuous scale rather than assigning music to a fixed set of discrete adjectives; we adopt this emotion model in this study.

Despite significant research efforts, quantifying emotion in music remains a complex challenge due to the subjectivity of human perception, which varies across individuals based on factors such as cultural background and personal preference. Over the years, researchers have invested significant time and effort into obtaining reliable emotion annotations. However, these are often collected under specific conditions, such as a focus on particular musical styles or genres \cite{Soleymani20131000SF,aljanaki2017deam,eerola_2019,Zhang2018ThePD,fan2020wcmed}. As a result, there are inherent limitations to their generalizability to other unseen musical styles. Previous studies have shown that different musical genres evoke distinct emotional responses \cite{eerola2011genre}, highlighting the challenge to the universality of emotion annotations across genres. This challenge becomes even more pronounced when considering multiple datasets where the annotation process was conducted under varying instructions and settings. The effect is particularly significant in audio-based models, as opposed to symbolic representations (e.g., MIDI), since audio retains all expressive qualities, including timbre, which strongly influences emotion perception \cite{wu2014timbre,hailstone2009timbre,panda2023feature}. Additionally, in dimensional MER, datasets are often compiled using different valence-arousal scales, making the integration challenging and possibly obscuring genre-specific emotional patterns.

In this work, we analyze the data distribution gap and genre bias in commonly used MER datasets. To investigate this, we start by determining a suitable input feature representation for the task of MER, which leads us to focus on Jukebox embeddings\footnote{The complementary code is available at \url{https://github.com/joann8512/emo-data-gap}.}. We demonstrate the data distribution gap with a systematic cross-dataset experiment. To better understand genre-emotion relationships in existing data, we analyze distribution divergences between datasets in both audio content and emotion annotations. Additionally, based on feature clustering, we uncover potential genre dominance and dataset biases in certain feature representations. Finally, we identify chroma features as a very simple complement to Jukebox embeddings that, combined with diversified training datasets, improves and stabilizes in- and out-of-distribution MER performance, enhancing adaptability to unseen data.

\section{Background}\label{sec:related_work}
Prior work in dimensional music emotion recognition has focused on identifying relevant audio features, drawing from psychological theories \cite{lu2006mooddetect}, music domain knowledge \cite{laurier2007mirex,yang2008regression,panda2011moodsvm}, and multi-modal sources such as MIDI and lyrics \cite{malheiro2018lyricemo,panda2013multi}. To examine the relationship between these features and emotion labels, tools such as PsySound \cite{cabrera1999psysound}, MARSYAS \cite{tzanetakis2002marsyas}, and MIRToolBox \cite{lartillot2007mirtoolbox} are often used \cite{aljanaki2013mireval,panda2020novel}. However, some of these are no longer maintained, making feature extraction and reproducibility increasingly difficult.

With the rise of deep audio embeddings such as VGGish \cite{Simonyan2014VeryDC} and OpenL3 \cite{cramer2019L3} in the MIR field, researchers have explored their use in MER tasks \cite{koh2021deepaudio}. More recently, large-scale models trained on massive datasets have shown strong performance and transferability across tasks \cite{li2023mert,won2024foundation,Gardner2023LLarkAM}. Castellon et al. \cite{Castellon2021CodifiedAL} demonstrated the effectiveness of the Jukebox embeddings \cite{dhariwal2020jukebox} in MER, outperforming other input types on EmoMusic, despite them not yet being commonly used in MIR tasks. Based on these results (and our own preliminary experiments), Jukebox embeddings will be used as the central feature set in our experiments (see Section \ref{sec:exp_setup}).

\section{Datasets}\label{sec:dataset}
As mentioned in Section \ref{sec:introduction}, we focus on datasets that provide dimensional annotations of perceived emotions, with the additional constraint that such remain publicly available. Among the widely used datasets since 2020 \cite{kang2024emosurvey}, EmoMusic, DEAM, and PMEmo are relatively well-established for MER tasks. The distribution of genre labels officially provided by EmoMusic and DEAM is plotted in Appendix \ref{sec:appendix_1} and documents a somewhat selective focus on popular genres (e.g. Country, Electronic, Jazz, and Pop).\footnote{Upon listening to example files, we found pieces labeled "Classical" in EmoMusic to be rather arbitrary items, including piano and drums, synthesized strings in the background, etc., so the "Classical" bars should be taken with a big grain of salt.} To investigate the performance on stylistically distant music, we will further include WTC and WCMED, both of which consist of Western classical piano excerpts, rather than merely expanding the corpus with large-scale commercial music datasets. Finally, for datasets that provide both static and dynamic annotations, only the static data is considered in our study. Below is a brief overview of the datasets:

\begin{enumerate}
  \item \textbf{EmoMusic (E)} \cite{Soleymani20131000SF} was developed for the "Emotion in Music" task at MediaEval 2013.\footnote{\url{http://www.multimediaeval.org/mediaeval2013/emotion2013/index.html}} The dataset consists of 744 audio recordings, each with a duration of 45 seconds, spanning a variety of musical styles, such as Country, Blues, Electronic, and Rock. The emotion annotations for the corresponding 45-second clips are on a continuous scale ranging from $-1$ to $+1$. 
  
  \item \textbf{DEAM (D)} \cite{aljanaki2017deam} extends the EmoMusic dataset, aggregating recordings from the MediaEval task between 2013 and 2015.\footnote{\url{http://www.multimediaeval.org/mediaeval2015/emotioninmusic2015/}} With a total of 1,802 tracks, it encompasses EmoMusic as a subset.
  
  \item \textbf{PMEmo (P)} \cite{Zhang2018ThePD} contains emotion annotations for 794 songs collected from the Billboard Hot 100, the iTunes Top 100 Songs (USA), and the UK Top 40 Singles Chart. Of these, 767 tracks include valence-arousal labels, annotated on manually selected chorus excerpts with values ranging from $0$ to $+1$. The duration of these excerpts varies from 11-88 seconds.
  
  \item \textbf{WTC (W1)} \cite{chowdhury2021wtc} provides valence-arousal annotations for six different performances, by six renowned pianists, of Bach’s \textit{Well-Tempered Clavier (WTC)} Book 1. The dataset maintains stylistic coherence, featuring compositions evenly distributed across all 24 possible major and minor keys, each represented by a prelude-fugue pair. In total, it contains 288 recordings, with participants rating valence on a scale of $-5$ to $+5$ and arousal on a scale of $0$ to $100$.

  \item \textbf{WCMED (W2)} \cite{fan2020wcmed} comprises 200 royalty-free audio recordings of Western classical repertoire, 79 of which are solo piano pieces by composers such as Bach, Beethoven, Chopin, Mozart, Rachmaninoff, collected from the Saarland Music Dataset (SMD) \cite{muller2011smd}. Each annotated excerpt ranges from 8 to 20 seconds in duration. Instead of absolute ratings, annotations were obtained through a ranking-based crowd-sourcing experiment, with participants performing pairwise comparisons. The resulting rankings span from $0$ to $400$ for both valence and arousal. While the investigation of adaptability in pairwise and direct annotations should be further studied, we include this set due to its representation of the Western classical music style.
\end{enumerate}

\section{Experiments}\label{sec:exp_setup}

In this section, we analyze the data distribution gap and genre bias in several steps, as outlined in the introduction. To establish a suitable audio feature set for the subsequent investigations, we evaluate various feature representations in an MER task on the MediaEval EmoMusic dataset, leading us to focus on Jukebox embeddings (Section \ref{sec:exp_features}). Using this feature set, we then demonstrate and quantify the data distribution gap with a cross-dataset experiment (Section \ref{sec:exp_crossexp}). Section \ref{sec:exp_divergence} analyzes distribution divergences between datasets in terms of both audio content and emotion annotations. Finally, we identify chroma features as a stabilizing factor and, in Section \ref{sec:exp_final} test how and to what extent the combination of Jukebox embeddings with chroma features improves in-distribution performance and out-of-distribution generalization. 

The underlying model architecture that we use for all subsequent learning experiments is a simple feedforward neural network (MLP) using Mean Squared Error (MSE) as the loss function. The input feature dimension is variable, depending on the type of input feature. The model consists of two hidden layers with ReLU activation, where the first hidden layer has 1024 units and the second has 512. To prevent overfitting, dropout is applied at the input and after each hidden layer. Finally, the model outputs two separate predictions, one for valence and one for arousal.

In terms of inputs, audio segment lengths vary across datasets (see Section \ref{sec:dataset} for details) to ensure that the emotion remains consistent throughout the annotated audio clip. During Jukebox embedding extraction, a random 25-second segment is selected from each annotated clip; segments shorter than this length are padded. As a result, audio clips of arbitrary lengths always produce a fixed-size input.

For all datasets considered in this work, the emotion annotations are independently normalized to the range of $-1$ and $+1$ by dataset. We follow the experimental setup of previous works and split each dataset into training, validation, and test sets using a ratio of $8:1:1$. This consistent split is applied across all experiments, whether using individual datasets or combining multiple. The exception is when a set is used as out-of-distribution test set, in which case the whole dataset is utilised.

\begin{table}[t]
\centering
\caption{Comparison of impact of input representations, trained and tested on EmoMusic. Results on test set are in terms of coefficient of determination ($R^2$). 
}
\begin{tabular}{l@{\hskip 10pt}c@{\hskip 10pt}c@{\hskip 10pt}c@{\hskip 10pt}}
\toprule
\textbf{Representation} & \textbf{Avg.} & \textbf{A.} & \textbf{V.} \\ \hline
Chroma             & 0.259    & 0.261    & 0.258 \\
MFCCs               & 0.499    & 0.579    & 0.419 \\
MidLevel\cite{Chowdhury2019TowardsEM}\tablefootnote{\url{https://github.com/shreyanc/midlevel\_general}}           & 0.384    & 0.387    & 0.382 \\
Encodec\cite{defossez2022highfi}\tablefootnote{\url{https://github.com/facebookresearch/encodec}}            & 0.448    & 0.527    & 0.370 \\
Music2Latent\cite{pasini2024musiclatent}\tablefootnote{\url{https://github.com/SonyCSLParis/music2latent}}       & 0.574    & 0.606    & 0.542 \\
MERT\cite{li2023mert}\tablefootnote{\url{https://github.com/yizhilll/MERT}}               & 0.614   & 0.656     & 0.572 \\ 
\textbf{Jukebox}   & \textbf{0.674}  & \textbf{0.708}  & \textbf{0.640} \\
\bottomrule
\end{tabular}
\label{tab:rep}
\end{table}

\subsection{Identifying the Best Audio Features}
\label{sec:exp_features}

\begin{table}[!t]
\centering
\caption{Cross-dataset evaluation with the Jukebox embeddings and a simple MLP with two hidden layers. Performances are compared using the coefficient of determination ($R^2$). Dataset names in the \textit{Test} column are abbreviated by their first letter, in the same order.}
\label{tab:cross-data}
\setlength{\tabcolsep}{2.2pt}
{\scriptsize
\begin{tabular}{l!{\vrule width 1.2pt} ccc !{\vrule width 1pt} ccc !{\vrule width 1pt} ccc !{\vrule width 1pt} ccc !{\vrule width 1pt} ccc}
\toprule
\multirow{2}{*}{\textbf{Test}} 
  & \multicolumn{3}{c!{\vrule width 1pt}}{\textbf{EmoMusic}} 
  & \multicolumn{3}{c!{\vrule width 1pt}}{\textbf{DEAM}} 
  & \multicolumn{3}{c!{\vrule width 1pt}}{\textbf{PMEmo}} 
  & \multicolumn{3}{c!{\vrule width 1pt}}{\textbf{WTC}} 
  & \multicolumn{3}{c}{\textbf{WCMED}} \\
\cmidrule{2-16}
   & Avg. & A. & V. 
   & Avg. & A. & V. 
   & Avg. & A. & V. 
   & Avg. & A. & V.
   & Avg. & A. & V. \\
\midrule
\textbf{E} 
    & \textbf{0.67} & \textbf{0.71} & \textbf{0.64} 
    & 0.44 & 0.53 & 0.36 
    & -0.32 & -0.55 & -0.09 
    & 0.05 & 0.23 & -0.13 
    & -0.26 & 0.16 & -0.68 \\
\textbf{D}  
    & 0.45 & 0.49 & 0.42 
    & \textbf{0.61} & \textbf{0.63} & \textbf{0.59} 
    & -0.45 & -0.51 & -0.38 
    & 0.14 & 0.29 & -0.01 
    & -0.17 & 0.06 & -0.40 \\
\textbf{P}  
    & 0.04 & -0.08 & 0.17 
    & -0.12 & 0.02 & -0.27 
    & \textbf{0.61} & \textbf{0.72} & \textbf{0.51} 
    & 0.02 & -0.02 & 0.05 
    & -0.62 & -0.24 & -1.01 \\
\textbf{W1}  
    & 0.29 & 0.22 & 0.36 
    & 0.45 & 0.61 & 0.29 
    & -0.62 & -1.56 & 0.33
    & \textbf{0.85} & \textbf{0.88} & \textbf{0.82} 
    & -0.15 & 0.58 & -0.88 \\
\textbf{W2}  
    & -0.84 & -1.12 & -0.56 
    & 0.04 & 0.18 & -0.10 
    & -0.62 & -0.92 & -0.31 
    & 0.17 & 0.23 & 0.11 
    & \textbf{0.81} & \textbf{0.75} & \textbf{0.87} \\
\bottomrule
\end{tabular}

}
\end{table}

Considering recent trends in learning representations with foundation models, we evaluate the effectiveness of Jukebox embeddings by comparing them against hand-crafted features (Chroma, MFCCs), data-learned features (Mid-Level Features), and embeddings from foundation models (MERT, Music2Latent, Encodec), using a consistent experimental setup as the previous step with minimal parameter tuning for each feature type (Table \ref{tab:rep}).

To ensure comparability, we follow the pipeline of Castellon et al.~\cite{Castellon2021CodifiedAL} to extract embeddings from the 36th (middle) layer of the Jukebox-5B model, yielding 4800-dimensional vectors. Performance is evaluated using the \textit{coefficient of determination} ($R^2$), a standard regression metric in MER to quantify how well the model explains variance in valence and arousal.
The hand-crafted features are computed using Librosa’s CQT-chromagram to capture tonal and harmonic content and Mel-frequency cepstral coefficients for the short-term spectral variations and temporal evolution. To enrich these features, we compute first-, second-, and third-order differences along the time axis, which are concatenated into a final feature vector. For data-driven approaches, Mid-level Features and foundation model embeddings are obtained using their respective open-source implementations (see references in Table \ref{tab:rep}). Results show that Jukebox embeddings outperform all alternative feature representations, and will thus serve as our primary feature set moving forward.

\subsection{The Gap: Cross-dataset Prediction}
\label{sec:exp_crossexp}
To assess the model's adaptability to unseen datasets, we conduct a series of systematic cross-dataset evaluations, using Jukebox embeddings as input features for all datasets, with a consistent model structure. Due to content similarities between EmoMusic and DEAM, training and testing interchangeably on these datasets yields relatively good results. However, performance drops significantly when evaluated on other datasets (see Table \ref{tab:cross-data}), which implies a severe data distribution gap. Since Jukebox is trained as a generative model capable of producing music with coherent tempo, genre, instrumentation, and key, its embeddings likely encode musically relevant information, making them effective in distinguishing patterns across datasets.

\begin{figure}
\centering
    \includegraphics[width=0.8\linewidth]{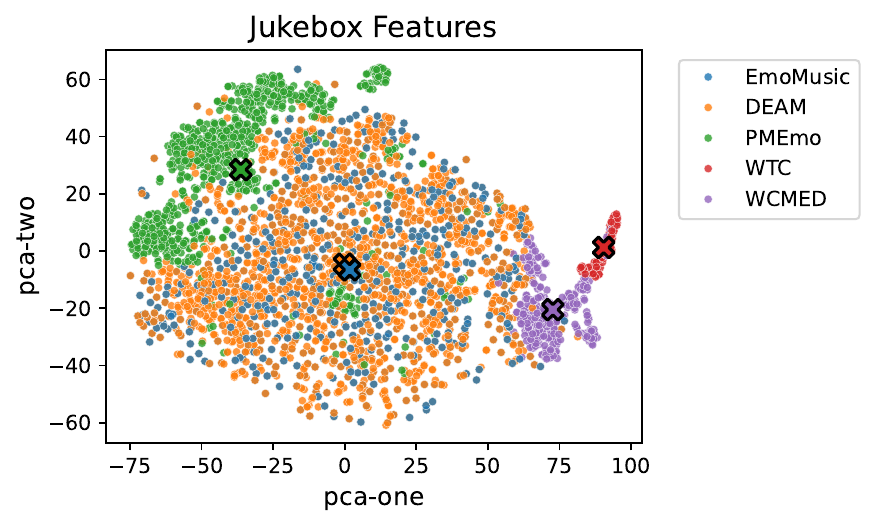}
\caption{Jukebox embeddings of each dataset visualized using t-SNE, with the $\times$ symbols indicating the centroids of each dataset. The close relation between DEAM and EmoMusic can clearly be seen. Note the extreme positions of the two classical music sets WTC and WCMED, which nevertheless still appear to be distinct from each other in terms of feature representation.}
\label{fig:tsne_jukebox}
\end{figure}

To further investigate, we apply t-SNE to the Jukebox embeddings from all datasets to visualize their distributions (Fig.~\ref{fig:tsne_jukebox}). The embeddings form distinct clusters by dataset, with the exception of EmoMusic and DEAM, which overlap due to shared content. The same visualization procedure is applied to all other feature types; all, except chroma, show similar dataset-wise separability (see Appendix \ref{sec:appendix_2}).

\subsection{Analyzing the Gap}
\label{sec:exp_divergence}

\renewcommand{\arraystretch}{1.1}
\newcommand{\smallnum}[1]{{\fontsize{8.5pt}{10pt}\selectfont #1}}
\begin{table*}[t] 
\centering
\caption{The Wasserstein distance (WD) and Jensen-Shannon divergence (JS) are calculated for both data content and emotion annotations across pairs of datasets. Each entry is shown in the format WD / JS. 
}

\begin{adjustbox}{width=\textwidth}
{\begin{tabular}{l|
    c@{\hskip 6pt}c|
    c@{\hskip 6pt}c|
    c@{\hskip 6pt}c|
    c@{\hskip 6pt}c}
\toprule
\multicolumn{1}{l|}{\textbf{}}     & \multicolumn{2}{c|}{\textbf{\smallnum{EmoMusic}}}     & \multicolumn{2}{c|}{\textbf{\smallnum{DEAM}}}     & \multicolumn{2}{c|}{\textbf{\smallnum{PMEmo}}}     & \multicolumn{2}{c}{\textbf{\smallnum{WTC}}}     \\ \hline
& \multicolumn{1}{c}{\smallnum{Data}\hphantom{.}}    & \smallnum{Annot.}    & \multicolumn{1}{c}{\smallnum{Data}\hphantom{.}}    & \smallnum{Annot.}    & \multicolumn{1}{c}{\smallnum{Data}\hphantom{.}}    & \smallnum{Annot.}    & \multicolumn{1}{c}{\smallnum{Data}\hphantom{.}}    & \smallnum{Annot.}     \\ \hline
\multicolumn{1}{l|}{\textbf{\smallnum{E}}} & \multicolumn{1}{c}{-} & -             & \multicolumn{1}{c}{-} & {-} & \multicolumn{1}{c}{-} & {-} & \multicolumn{1}{c}{-} & {-} \\
\multicolumn{1}{l|}{\textbf{\smallnum{D}}}     & \multicolumn{1}{c}{\smallnum{0.03/0.02}\hphantom{.}}            & \smallnum{0.05/0.13} & \multicolumn{1}{c}{-}                        & -                        & \multicolumn{1}{c}{-}                        & -                      & \multicolumn{1}{c}{-}                                    & -             \\
\multicolumn{1}{l|}{\textbf{\smallnum{P}}}    & \multicolumn{1}{c}{\smallnum{0.20/0.15}\hphantom{.}}            & \smallnum{0.11/0.03}  & \multicolumn{1}{c}{\smallnum{0.19/0.14}\hphantom{.}}            & \smallnum{0.10/0.16}            & \multicolumn{1}{c}{-}                        & -                        & \multicolumn{1}{c}{-}      & - \\
\multicolumn{1}{l|}{\textbf{\smallnum{W1}}}      & \multicolumn{1}{c}{\smallnum{0.37/0.25}\hphantom{.}}            & \smallnum{0.12/0.49} & \multicolumn{1}{c}{\smallnum{0.45/0.31}\hphantom{.}}            & \smallnum{0.18/0.47}            & \multicolumn{1}{c}{\smallnum{0.37/0.26}\hphantom{.}}            & \smallnum{0.15/0.60}            &  
\multicolumn{1}{c}{-}                                    & -  \\
\multicolumn{1}{l|}{\textbf{\smallnum{W2}}}    & \multicolumn{1}{c}{\smallnum{1.71/0.46}\hphantom{.}}            & \smallnum{0.14/0.02} & \multicolumn{1}{c}{\smallnum{1.74/0.47}\hphantom{.}}            & \smallnum{0.19/0.05}            & \multicolumn{1}{c}{\smallnum{1.71/0.46}\hphantom{.}}            & \smallnum{0.15/0.11}            &        
\multicolumn{1}{c}{{\smallnum{1.59/0.43}\hphantom{.}}} & \smallnum{0.13/0.51}\\
\bottomrule
\end{tabular}}
\end{adjustbox}
\label{tab:data_dist}
\end{table*}

The universality of emotion annotations across genres is particularly important in our study, as we aim to investigate the potential of integrating annotations from multiple datasets for MER. However, as different genres can evoke distinct emotional responses \cite{eerola2011genre}, this raises concerns about annotation consistency across datasets. Because these annotations are independently normalized -- due to varying annotation scales across datasets -- this process may obscure differences in the underlying data distributions. Consequently, there may be a mismatch between the musical content and the seemingly aligned emotion annotations. To examine this relationship, we compute the Wasserstein distance and Jensen-Shannon divergence between all dataset pairs, comparing both the data content distributions and the emotion annotation distributions.
The Wasserstein distance quantifies distributional shifts by measuring the optimal transport cost needed to align two embedding sets. The JS divergence, a symmetrized and smoothed version of the Kullback-Leibler divergence, evaluates the similarity in statistical properties between distributions.

\begin{figure*}[t]
\centering
    \begin{subfigure}[b]{\textwidth}
      \centering
      \includegraphics[width=1.0\textwidth]{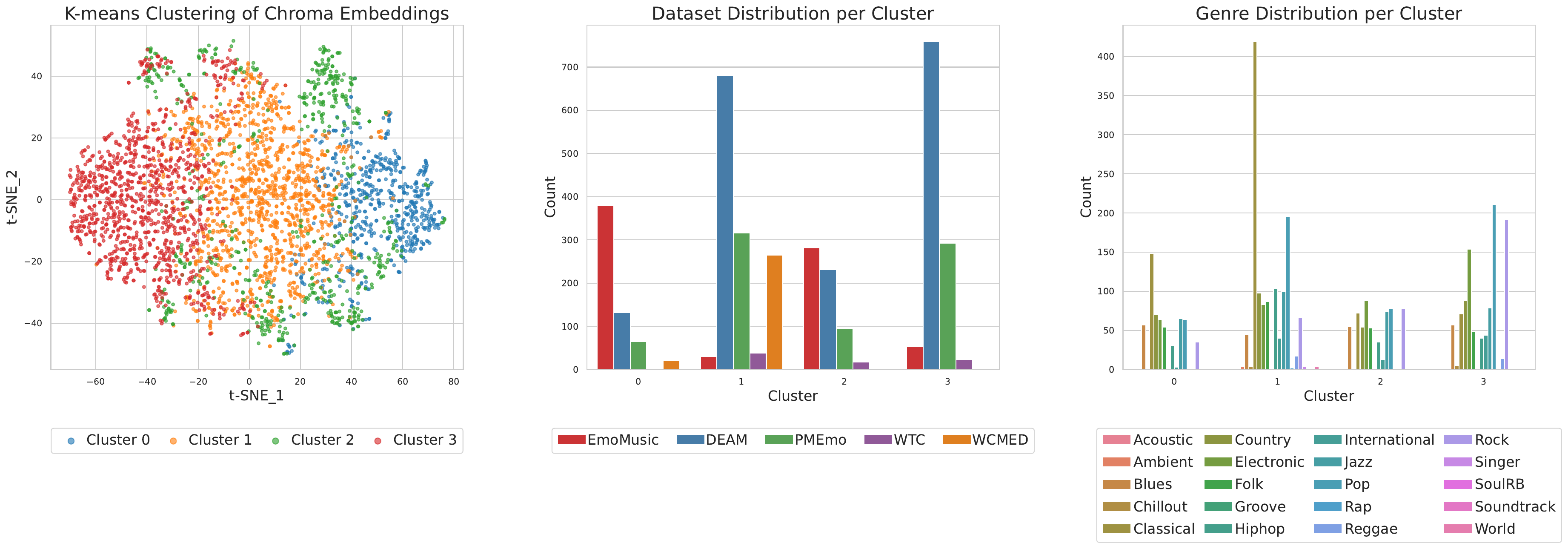}
      \includegraphics[width=1.0\textwidth]{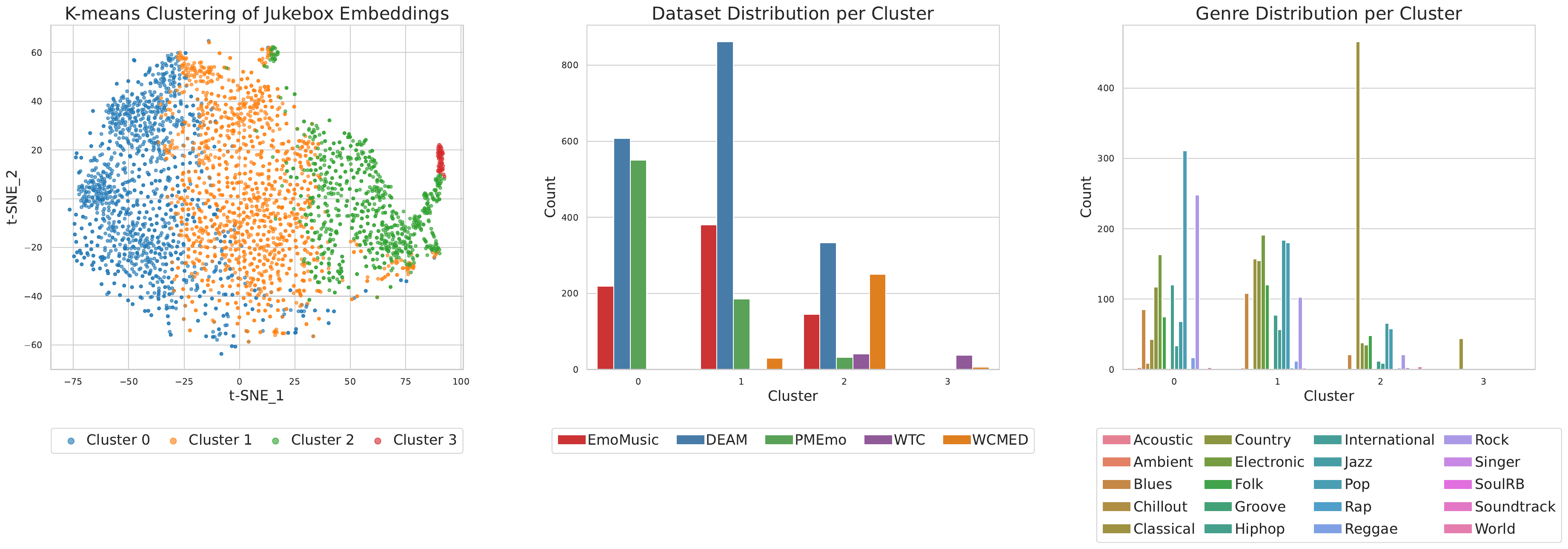}
      \label{fig:chroma_kmean}
    \end{subfigure}%
\put(-290,-4){\makebox(0,0){\small (a)}}
\put(-175,-4){\makebox(0,0){\small (b)}}
\put(-55,-4){\makebox(0,0){\small (c)}}
\caption{Visualization of K-means clustering on Chroma (top) and Jukebox (bottom) features, along with the corresponding dataset and genre distribution.
}
\label{fig:cluster_plots}
\end{figure*}

The results are shown in Table \ref{tab:data_dist}.
To interpret the pairwise comparisons, consider the EmoMusic column as an example. When compared to DEAM, the data distributions (column ``Data'') are highly similar (WD=0.03, JS=0.02) but the annotation distributions (column ``Annotation'') show a slightly larger divergence (WD=0.05, JS=0.13), suggesting differences in emotion interpretation. On the other hand, when compared to PMEmo, data distributions exhibit a relatively small shift (WD=0.20, JS=0.15), while the annotation distributions indicate a stronger similarity in emotion labels (WD=0.11, JS=0.03). 

Interestingly (still in the EmoMusic column), comparison with the two classical music datasets reveals notable differences. The largest shift in data distribution, indicating a substantial difference in content, occurred between EmoMusic and WCMED (WD=1.71, JS=0.46), while the annotations remained relatively close (WD=0.14, JS=0.02). Despite the pronounced content differences, the distribution of emotion labels across these datasets seems to be strikingly similar. WTC, on the other hand, though coming under the same genre label as WCMED ("Classical"; and indeed, both datasets contain solo piano music only, but from different composers) shows only a moderate shift in data distribution to EmoMusic (WD=0.37, JS=0.25), but a notable difference in the interpretation of emotions, as shown by the notable difference in the emotion annotations (WD=0.12, JS=0.49). 
These findings not only highlight the differing associations of classical music datasets with EmoMusic but also suggest that even when music content varies profoundly, they may still point toward similar emotional representations.

To examine the factors contributing to the separation of Jukebox embeddings by dataset (Fig. \ref{fig:tsne_jukebox}), we apply k-means clustering to the Jukebox feature vectors to see if there are associations between certain features and certain datasets or genres. Motivated by preliminary findings suggesting that Chroma features have a stabilizing effect across datasets, we include Chroma features in the same clustering analysis. Indeed, it is quite reasonable to assume that Chroma features might be generally informative, as the perception of emotion in music is often linked to harmonic content, also in the literature.
More precisely, we perform k-means clustering on the combined Jukebox and Chroma feature vectors extracted from the union of all datasets, and analyze how strongly each dataset and genre is represented in each resulting cluster. The lower half of Fig.~\ref{fig:cluster_plots}(b) and (c) suggests that the Jukebox embeddings capture genre-specific characteristics, which is further supported by visual similarity between the clustering in Fig.~\ref{fig:tsne_jukebox}. In contrast, clusters formed using Chroma features show relatively uniform distributions over datasets and genres, indicating lower sensitivity to genre-specific differences. These findings suggest that combining Jukebox embeddings with Chroma features may mitigate dataset and genre bias in MER model training.


\begin{table*}[t] 
\centering
\caption{Results on in- and out-of-distribution emotion recognition. The first two lines report the results of \cite{Soleymani20131000SF} and \cite{koh2021deepaudio}; we only cite the corresponding numbers from these papers and did not reproduce their results. Testing with out-of-distribution datasets is only performed in our own experiments, from line 3 onwards. The \textit{combined} dataset is the union of EmoMusic, PMEmo, and WTC.}
\setlength{\tabcolsep}{7.0pt}
\begin{tabular}{l|l|l|ccc}
\toprule
\textbf{Input}                & \textbf{Training}        & \textbf{Testing} & \textbf{Avg.} & \textbf{A.} & \textbf{V.} \\ \hline
Hand-crafted \cite{Soleymani20131000SF}  & EmoMusic   & EmoMusic   & -   & 0.54   & 0.07 \\ \hline
OpenL3 \cite{koh2021deepaudio}  & EmoMusic   & EmoMusic   & -   & 0.67   & 0.56 \\ \hline
\multirow[b]{3.5}{*}{Jukebox}    & \multirow[b]{2}{*}{EmoMusic}       & EmoMusic    & 0.674    & 0.708    & 0.640    \\
& & DEAM      & 0.454    & 0.490    & 0.418    \\
& & WCMED     & -0.835   & -1.115   & -0.555   \\ \cline{2-6}
& \multirow[b]{2}{*}{Combined}       & Combined    & 0.632    & 0.685    & 0.580    \\
& & DEAM      & 0.619    & 0.618    & 0.620    \\
& & WCMED     & 0.082   & 0.336   & -0.172   \\ \hline
\multirow[b]{3.5}{*}{Jukebox$+$Chroma} & \multirow[b]{2}{*}{EmoMusic}  & EmoMusic    & 0.651    & 0.692    & 0.610    \\
& & DEAM      & 0.479    & 0.528    & 0.430.   \\
& & WCMED     & 0.002    & 0.232    & -0.228   \\ \cline{2-6}
& \multirow[b]{2}{*}{Combined} & Combined   
& 0.684  & 0.745  & 0.622 \\
& & DEAM     & 0.830     & 0.826    & 0.835    \\
& & WCMED    & 0.277     & 0.366    & 0.188    \\  
\bottomrule
\end{tabular}
\label{tab:final_table}
\end{table*}

\subsection{Bridging the Gap?}
\label{sec:exp_final}

As the final step in bridging the gap between datasets and genres, we revisit and combine the previous findings. Generally, we wish to introduce and combine data from a broader spectrum of musical styles, to train MER models that generalize better. But instead of combining all five datasets, we focus on three representing distinct styles: EmoMusic, PMEmo, and WTC. This allows us to perform out-of-distribution testing using the remaining two datasets. In addition, based on the findings presented above, and confirmed also relative to other feature sets in Fig.~4 in the Appendix \ref{sec:appendix_2}, we add the Chroma features to the input representation. As per our clustering experiment, they seem to be less dataset- and genre-specific and might have a regularizing effect. More precisely, we concatenate the 4800-dimensional Jukebox input vector with a 72-dimension vector derived from the 12 Chroma values, including the mean and standard deviation of their first- and second-order derivatives over time. 

As shown in Table \ref{tab:final_table}, adding Chroma features alone already substantially improves generalization to the out-of-domain Classical dataset WCMED, albeit at the cost of a slight drop in performance on the in-domain EmoMusic test set. Broadening the stylistic coverage of the training data by including PMEmo and WTC yields additional improvement, lifting the results on both the now more diverse in-domain test set (EmoMusic+PMEmo+WTC) and -- very substantially -- on the out-of-domain WCMED collection. DEAM, which is closely related to EmoMusic, also benefits from Chroma features and gains further improvement with a diversified training set.

To contextualize our results, the first two rows of Table \ref{tab:final_table} report scores from \cite{Soleymani20131000SF} and \cite{koh2021deepaudio}, both of which trained and tested their models on EmoMusic using the same $R^2$ metric for valence and arousal, making their results directly comparable to ours. The prior state of the art on this dataset would be \cite{Castellon2021CodifiedAL}, who first applied Jukebox embeddings to MER tasks. In our replication of their baseline setup (row 3 of Table~\ref{tab:final_table}), we obtain $R^2$ scores of 0.674, 0.708, 0.640. Furthermore, to isolate the impact of Chroma features, we conducted an experiment where only the training set was expanded, without including Chroma features (rows 6-8 of Table~\ref{tab:final_table}). While this setup already improves performance on out-of-distribution datasets compared to using only EmoMusic for training, the gains are notably smaller than those achieved by incorporating Chroma features.

\section{Conclusion}
\label{sec:conclusion}

With this project, we hope to have heightened the awareness, in the MER research community, of the problem of style- and genre-specificity of current MER models. We documented the problem with a systematic cross-dataset prediction experiment. Based on a series of experiments and analyses, we finally arrived at the combination of Jukebox embeddings and Chroma features as a simple, but apparently robust baseline representation which, in combination with a diversification of training data, permits us to train models that generalize substantially better to out-of-distribution data. We propose this representation and the results of the last experiment as a new baseline for the MER community.

\begin{credits}
\subsubsection{\ackname} This work was supported by the European Research Council (ERC) under the EU's Horizon 2020 research \& innovation programme, grant agreement No.\ 101019375 (\textit{Whither Music?}). 
The LIT AI Lab is funded by the Federal State of Upper Austria.
\end{credits}


\appendix
\section{Appendix}
\label{sec:appendix_1}

\begin{figure}[h]
    \centering
    \includegraphics[width=0.6\textwidth]{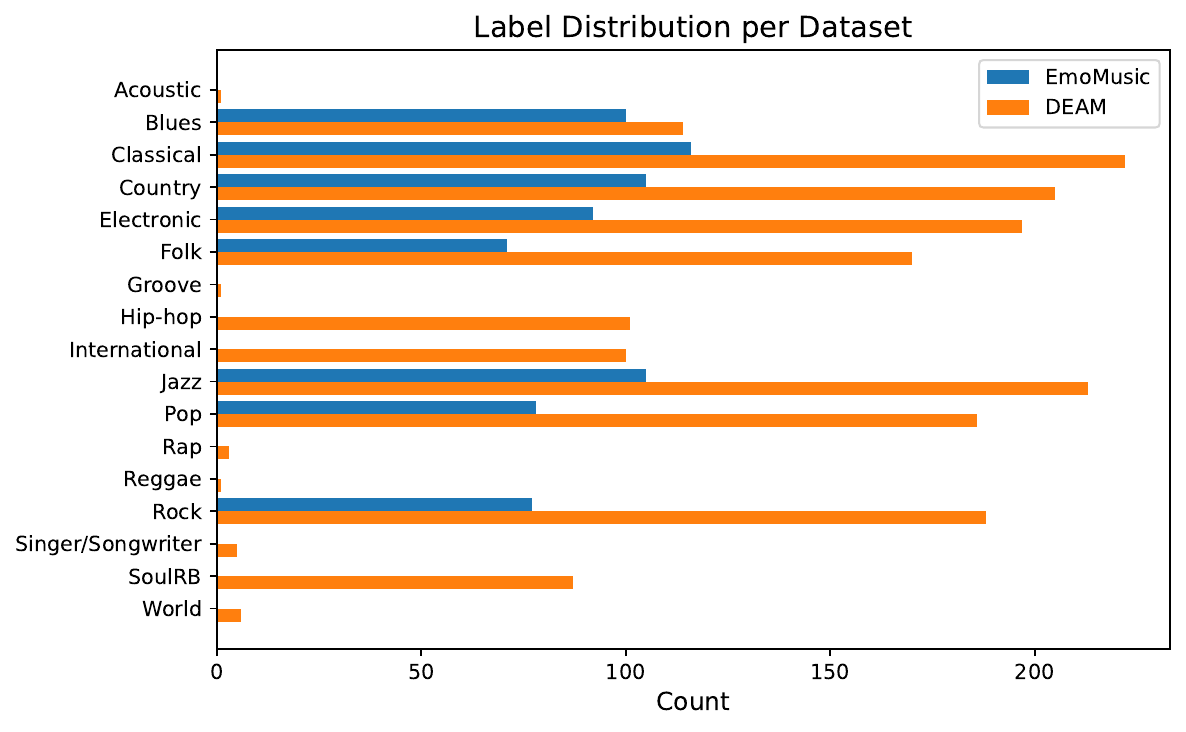}
    \caption[]{Genre distributions of EmoMusic and DEAM with the officially provided genre labels. Some genres are only present in DEAM, as EmoMusic is a smaller dataset and does not cover as broad a range of labels.}
    \label{fig:genre_plots}
\end{figure}

Figure~\ref{fig:genre_plots} in this document provides the genre distributions for the datasets EmoMusic and DEAM, as officially reported. While both datasets contain a broad variety of genres, they are predominantly focused on Country, Electronic, Jazz, and Pop. Although "Classical" appears to have the largest genre representation, we found pieces labeled as "Classical" to be somewhat arbitrary. For this reason, we expand the corpus for the experiments to include datasets consisting exclusively of Western classical music.

\begin{figure}[!htbp]
  \centering
  \begin{subfigure}[b]{0.45\textwidth}
    \centering
    \includegraphics[width=\linewidth]{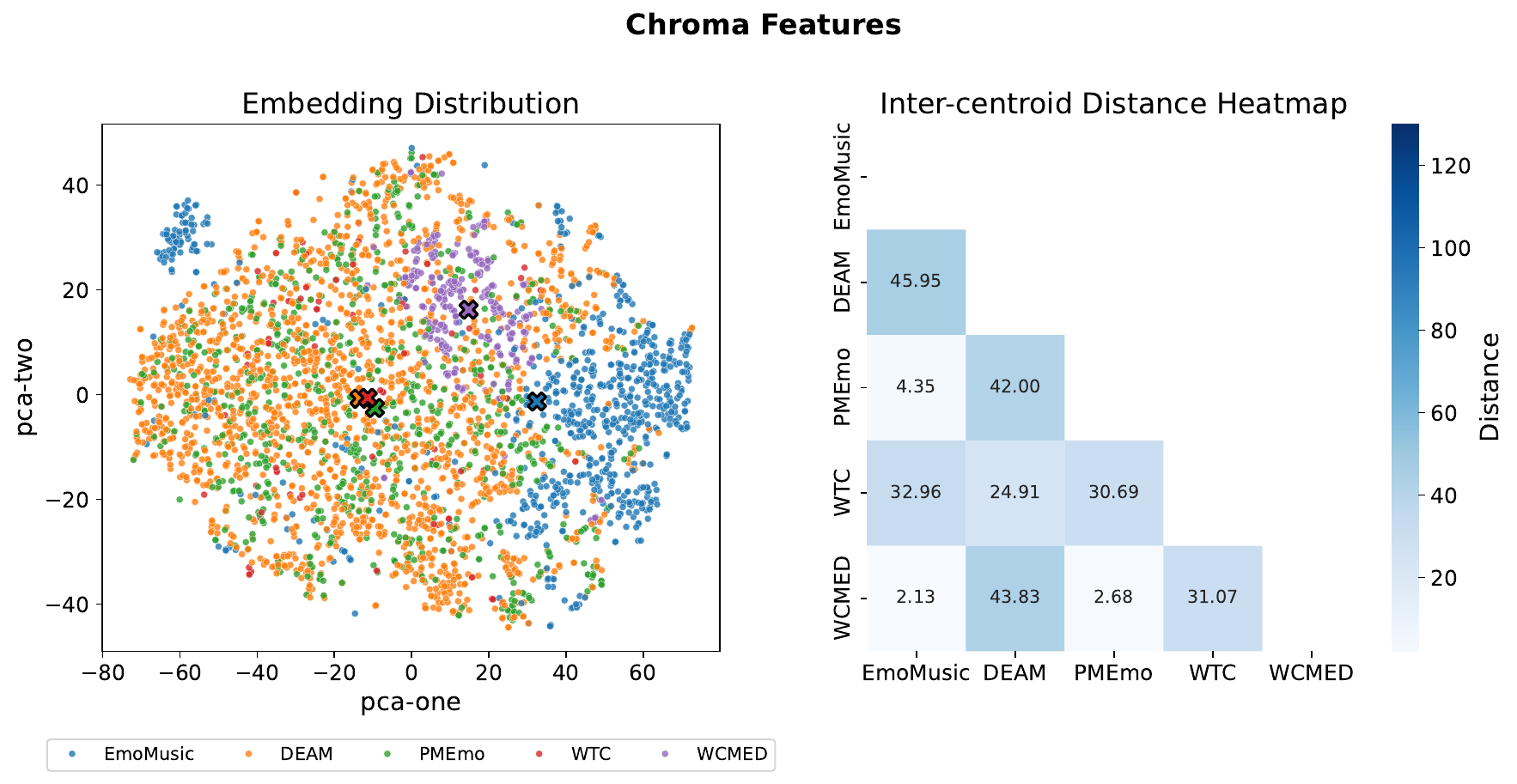}
    \captionsetup{justification=centering}
    \caption{\mbox{Inter-centroid Dist. (\textmu=28.23, \sigmasq=255.39)}}
  \end{subfigure}
  \hfill
  \begin{subfigure}[b]{0.45\textwidth}
    \centering
    \includegraphics[width=\linewidth]{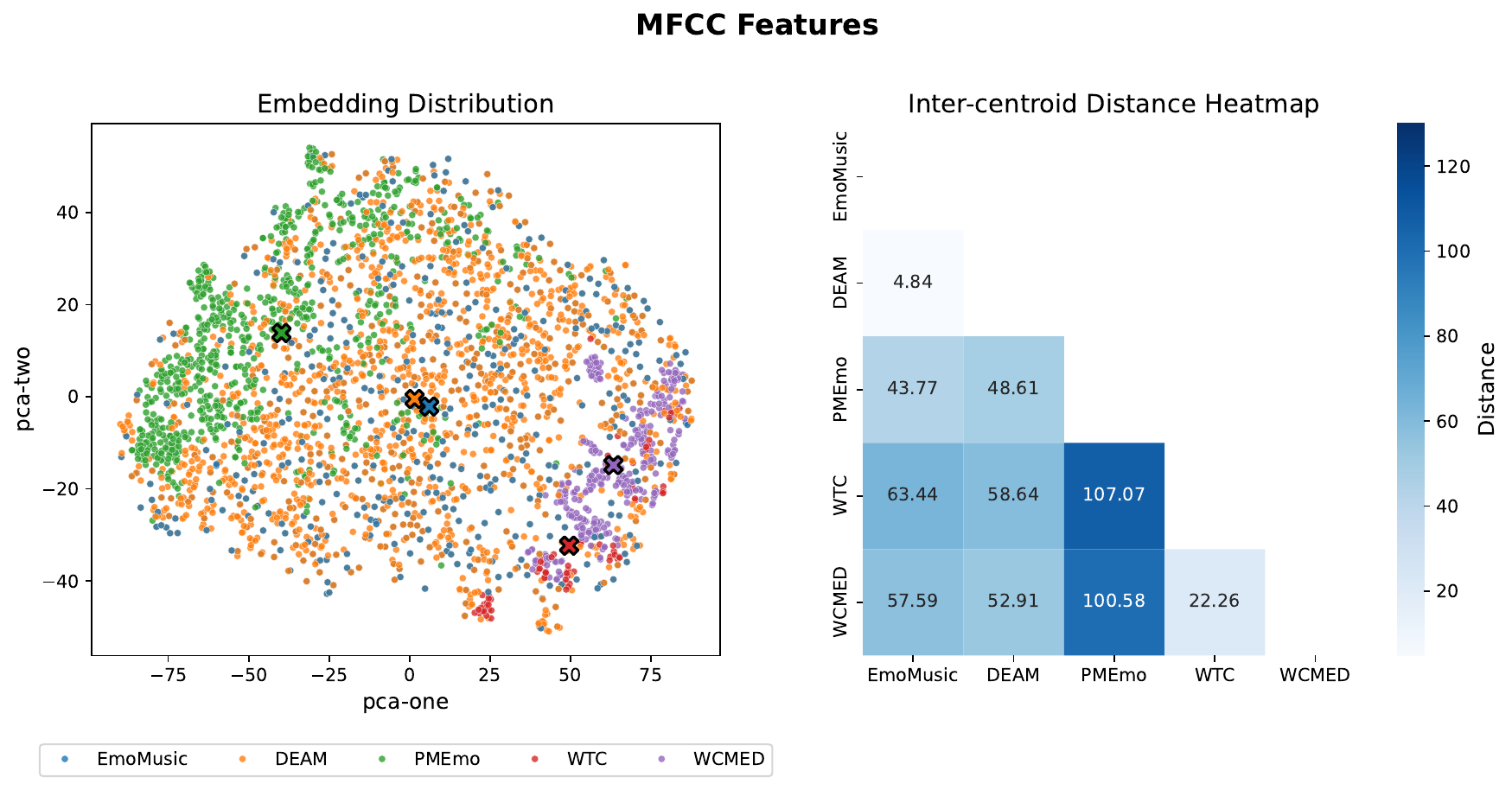}
    \captionsetup{justification=centering}
    \caption{\mbox{Inter-centroid Dist. (\textmu=55.97, \sigmasq=862.96)}}
  \end{subfigure}

  \vspace{1em}

  \begin{subfigure}[b]{0.45\textwidth}
  \centering
    \includegraphics[width=\linewidth]{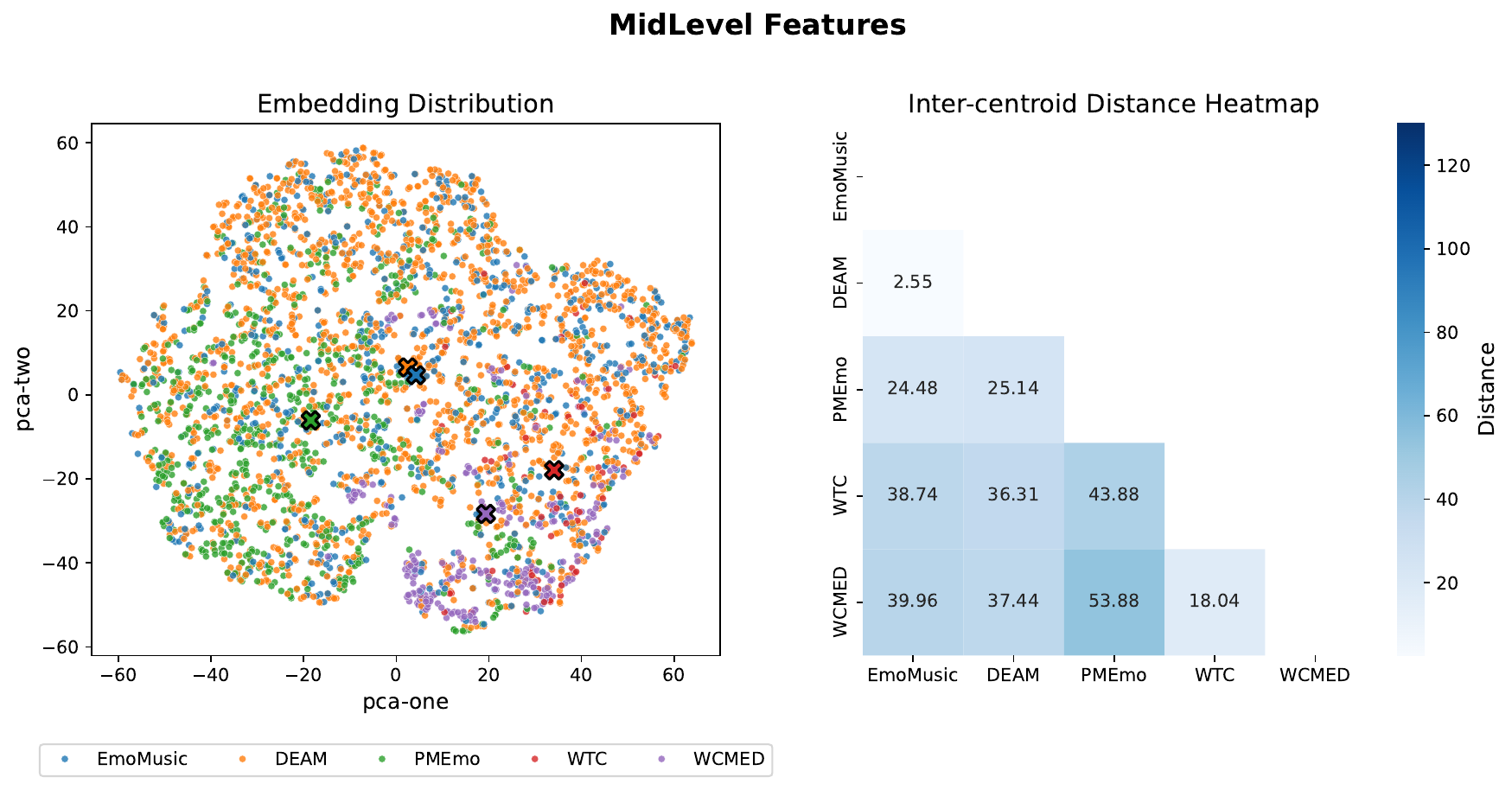}
    \captionsetup{justification=centering}
    \caption{\mbox{Inter-centroid Dist. (\textmu=32.04, \sigmasq=194.23)}}
  \end{subfigure}
  \hfill
  \begin{subfigure}[b]{0.45\textwidth}
  \centering
    \includegraphics[width=\linewidth]{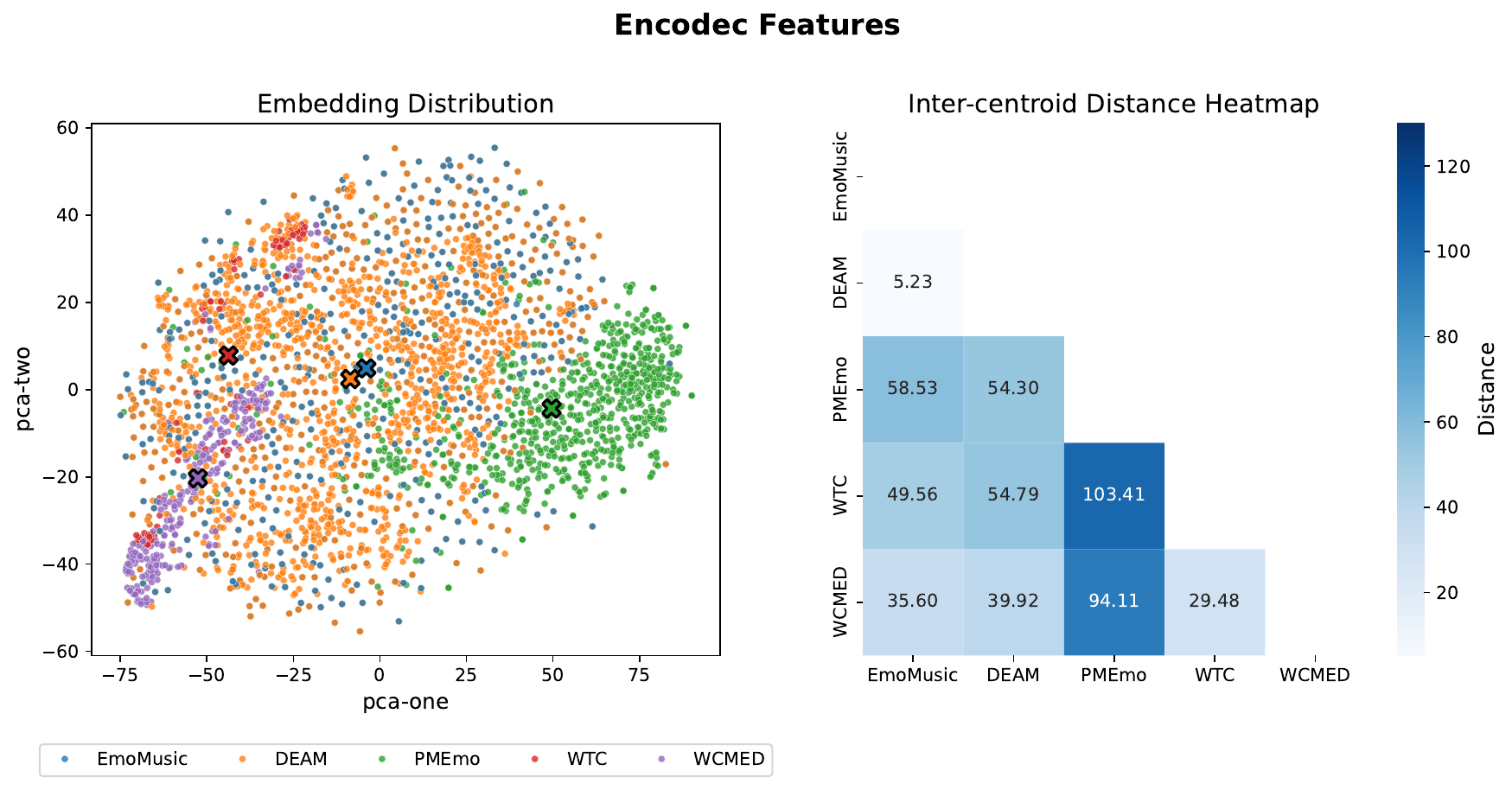}
    \captionsetup{justification=centering}
    \caption{\mbox{Inter-centroid Dist. (\textmu=52.49, \sigmasq=758.53)}}
  \end{subfigure}

  \vspace{1em}

  \begin{subfigure}[b]{0.45\textwidth}
  \centering
    \includegraphics[width=\linewidth]{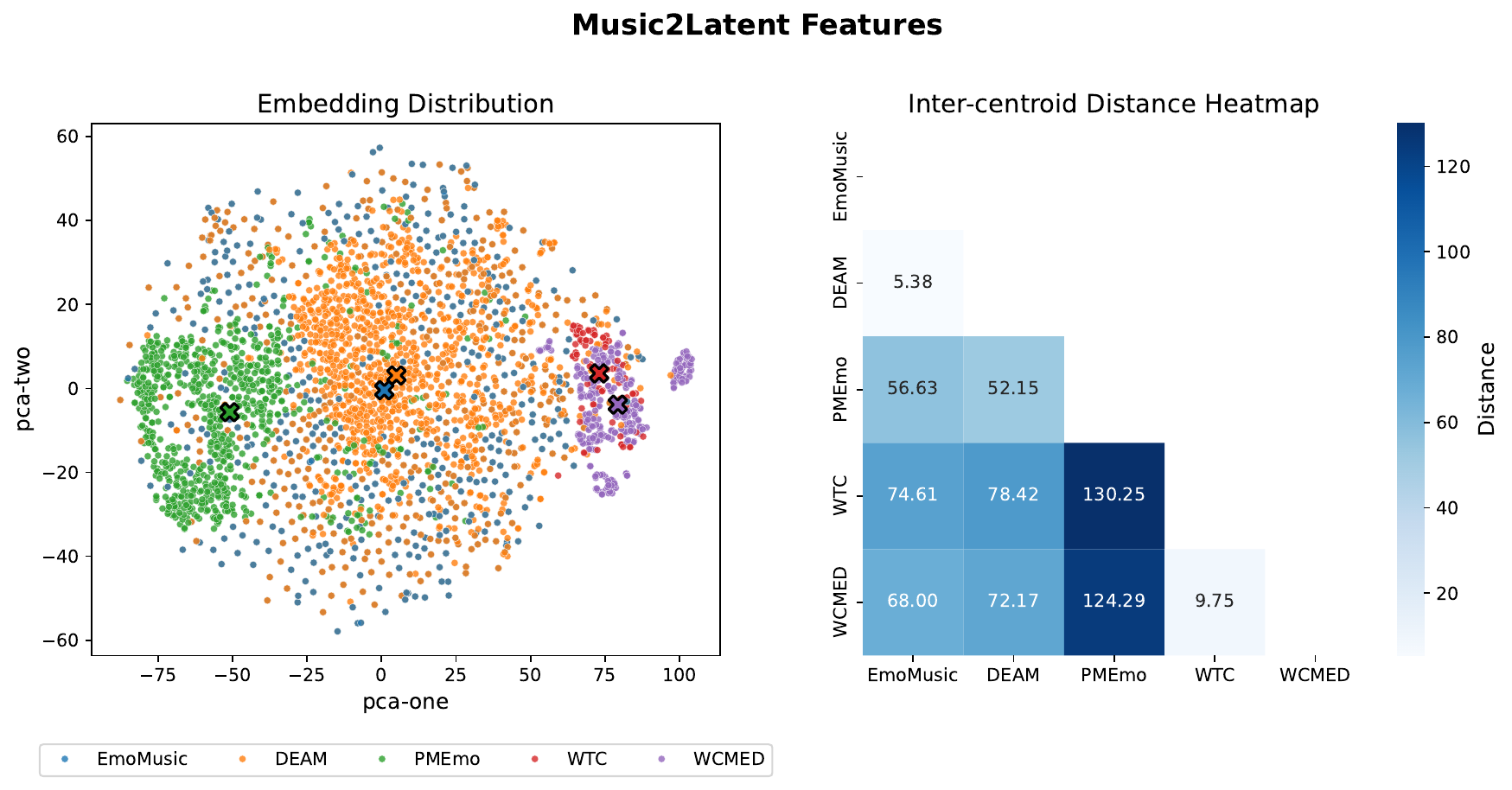}
    \captionsetup{justification=centering}
    \caption{\mbox{Inter-centroid Dist. (\textmu=67.16, \sigmasq=1490.04)}}
  \end{subfigure}
  \hfill
  \begin{subfigure}[b]{0.45\textwidth}
  \centering
    \includegraphics[width=\linewidth]{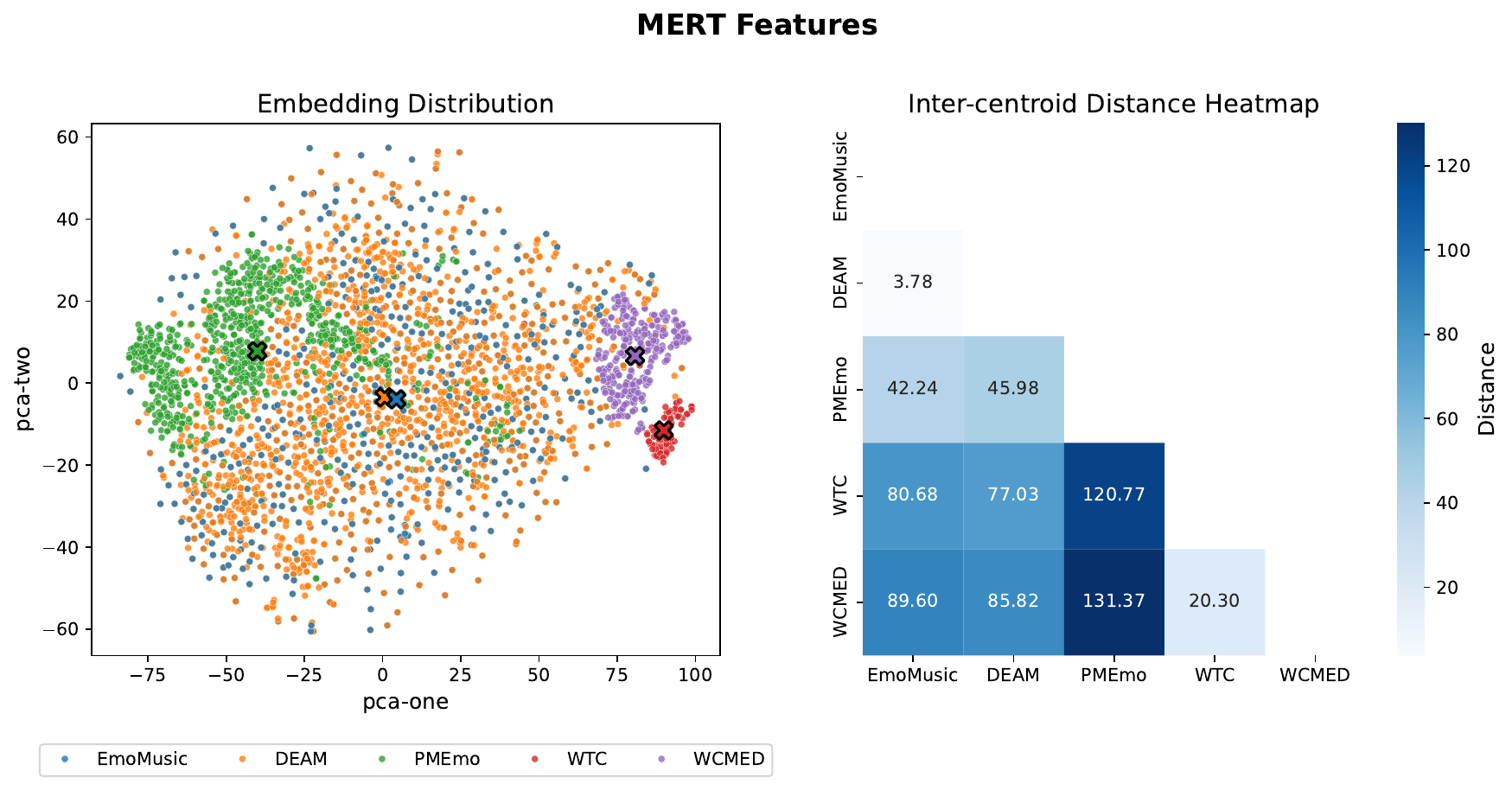}
    \captionsetup{justification=centering}
    \caption{\mbox{Inter-centroid Dist. (\textmu=69.76, \sigmasq=1534.43)}}
  \end{subfigure}

  \vspace{1em}
  
  \begin{subfigure}[b]{\textwidth}
  \centering
    \includegraphics[width=0.45\textwidth]{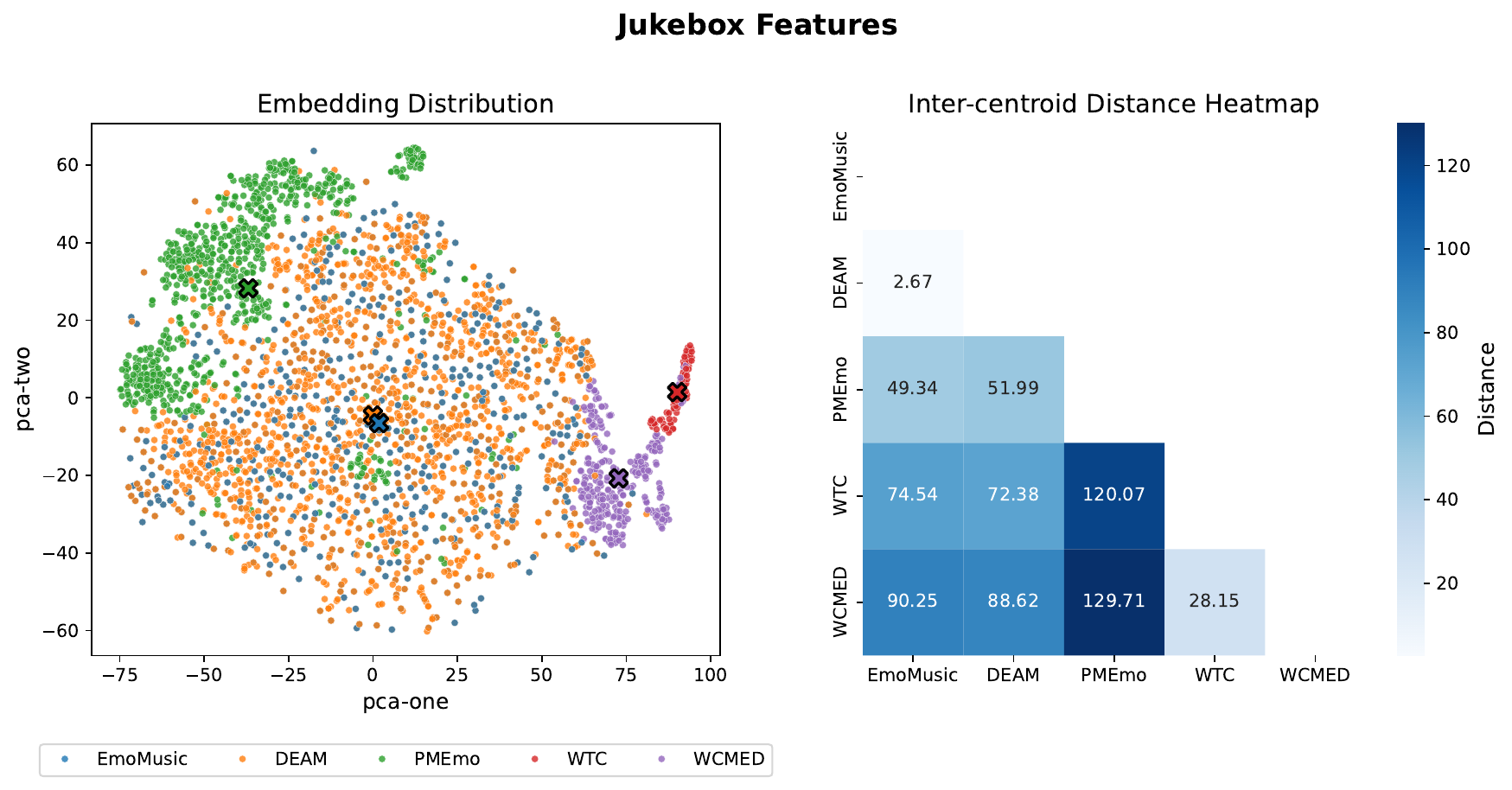}
    \caption{Inter-centroid Dist. (\textmu=70.77, \sigmasq=1388.54)}
  \end{subfigure}
  \caption{t-SNE visualization of all embeddings considered in the task, along with the corresponding inter-centroid distance heatmap for each dataset pair. The mean and variance of the inter-centroid distances are also reported.}
  \label{fig:feature_cluster}
\end{figure}

As mentioned in Section 4.2 of the main text, a clear pattern emerges when visualizing the embedding distributions with t-SNE: the embeddings are well-separated by dataset. Since Jukebox is a generative model trained to produce songs with consistent tempo, genre, instrumentation, and key, its embeddings likely capture musically relevant information, such as tonal, rhythmic, and timbral properties. To examine whether such distinctive clustering patterns are unique to Jukebox or also present in other embeddings without these properties, we apply the same visualization method to all other feature types considered in this study (Fig.~\ref{fig:feature_cluster}). Additionally, we calculate the \textit{inter-centroid distance}, as well as the average and variance, as a metric to compare the spread and relationships between datasets. These results are visualized as heatmaps, which show the distance between each dataset pair. Ideally, if the feature is not strongly distinguishable by dataset, we would expect a lower mean inter-centroid distance and variance. The plots reveal that all features, with the exception of Chroma features, exhibit a similar pattern: PMEmo predominantly occupies one side of the 2D plane, the two Western classical sets dominate the other side, and EmoMusic closely overlaps with DEAM due to their shared content. The inter-centroid distances between dataset pairs for all features also showed the same pattern. Therefore, suggesting that the bias observed in the Jukebox embeddings is not unique to that feature set.

\section{Appendix}
\label{sec:appendix_2}

%
%
%
\bibliographystyle{splncs04}
\bibliography{cmmr2025_paper}

\begin{thebibliography}{10}
\providecommand{\url}[1]{\texttt{#1}}
\providecommand{\urlprefix}{URL }
\providecommand{\doi}[1]{https://doi.org/#1}

\bibitem{aljanaki2013mireval}
Aljanaki, A., Wiering, F., Veltkamp, R.: {MIRU}trecht participation in mediaeval 2013: Emotion in music task. In: MultiMediaeval Benchmark Proceedings (2013)

\bibitem{aljanaki2017deam}
Aljanaki, A., Yang, Y.H., Soleymani, M.: Developing a benchmark for emotional analysis of music. PLOS ONE  \textbf{12},  e0173392 (2017)

\bibitem{cabrera1999psysound}
Cabrera, D.: Psysound: A computer program for psychoacoustical analysis. In: Australian Acoustical Society Conference (1999)

\bibitem{Castellon2021CodifiedAL}
Castellon, R., Donahue, C., Liang, P.: Codified audio language modeling learns useful representations for music information retrieval. In: Proc. 22nd International Society for Music Information Retrieval Conference (ISMIR) (2021)

\bibitem{cheuk2020triplet}
Cheuk, K.W., et~al.: Regression-based music emotion prediction using triplet neural networks. International Joint Conference on Neural Networks  (2020)

\bibitem{Chowdhury2019TowardsEM}
Chowdhury, S., Vall, A., Haunschmid, V., Widmer, G.: Towards explainable music emotion recognition: The route via mid-level features. In: Proc. 20th International Society for Music Information Retrieval Conference (ISMIR) (2019)

\bibitem{chowdhury2021wtc}
Chowdhury, S., Widmer, G.: On perceived emotion in expressive piano performance: Further experimental evidence for the relevance of mid-level perceptual features. In: Proc. 22nd International Society for Music Information Retrieval Conference (ISMIR) (2021)

\bibitem{cramer2019L3}
Cramer, J., Wu, H.H., Salamon, J., Bello, J.P.: Look, listen and learn more: Design choices for deep audio embeddings. In: Proc. IEEE Int.~Conf.~on Acoustics, Speech and Signal Processing (ICASSP) (2019)

\bibitem{dhariwal2020jukebox}
Dhariwal, P., et~al.: Jukebox: A generative model for music. arXiv preprint arXiv:2005.00341  (2020)

\bibitem{defossez2022highfi}
Défossez, A., Copet, J., Synnaeve, G., Adi, Y.: High fidelity neural audio compression. arXiv preprint arXiv:2210.13438  (2022)

\bibitem{eerola2011genre}
Eerola, T.: Are the emotions expressed in music genre-specific? {A}n audio-based evaluation of datasets spanning classical, film, pop and mixed genres. Journal of New Music Research  (2011)

\bibitem{eerola_2019}
Eerola, T.: Music and emotion stimulus sets consisting of film soundtracks. OSF (Aug 2019)

\bibitem{ekman1999basic}
Ekman, P.: Basic emotions. in Handbook of Cognition and Emotion, 1st ed. (1999)

\bibitem{fan2020wcmed}
Fan, J., Yang, Y.H., Dong, K., Pasquier, P.: A comparative study of western and chinese classical music based on soundscape models. In: IEEE Int.~Conf.~ on Acoustics, Speech and Signal Processing (ICASSP) (2020)

\bibitem{gabrielsson2002perceive}
Gabrielsson, A.: Emotion perceived and emotion felt: Same or different? Musicae Scientiae  \textbf{5},  123--147 (2001)

\bibitem{Gardner2023LLarkAM}
Gardner, J., Durand, S., Stoller, D., Bittner, R.M.: {LL}ark: A multimodal instruction-following language model for music. In: International Conference on Machine Learning (2023)

\bibitem{hailstone2009timbre}
Hailstone, J., et~al.: It's not what you play, it's how you play it: Timbre affects perception of emotion in music. Quarterly journal of experimental psychology  \textbf{62},  2141--55 (2009)

\bibitem{kang2024emosurvey}
Kang, J., Herremans, D.: Are we there yet? a brief survey of music emotion prediction datasets, models and outstanding challenges. arXiv preprint arXiv:2406.08809  (2024)

\bibitem{koh2021deepaudio}
Koh, E.S., Dubnov, S.: Comparison and analysis of deep audio embeddings for music emotion recognition. arXiv preprint arXiv:2104.06517  (2021)

\bibitem{kreutz2007induce}
Kreutz, G., Ott, U., Teichmann, D., Osawa, P., Vaitl, D.: Using music to induce emotions: Influences of musical preference and absorption. Psychology of Music  \textbf{36},  101--126 (2007)

\bibitem{lartillot2007mirtoolbox}
Lartillot, O., Toiviainen, P.: A matlab toolbox for musical feature extraction from audio. In: Proc. of the 10th International Conference of Digital Audio Effects (DAFx). p. 237–244 (2007)

\bibitem{laurier2007mirex}
Laurier, C., Herrera, P.: Audio music mood classification using support vector machine. MIREX Task on Audio Mood Classification  (2007)

\bibitem{li2023mert}
Li, Y., et~al.: {MERT}: Acoustic music understanding model with large-scale self-supervised training. International Conference on Learning Representations (ICLR)  (2023)

\bibitem{lu2006mooddetect}
Lu, L., Liu, D., Zhang, H.J.: Automatic mood detection and tracking of music audio signals. IEEE Transactions on Audio, Speech, and Language Processing  \textbf{14}(1),  5--18 (2006)

\bibitem{malheiro2018lyricemo}
Malheiro, R., Panda, R., Gomes, P., Paiva, R.P.: Emotionally-relevant features for classification and regression of music lyrics. IEEE Transactions on Affective Computing  \textbf{9}(2),  240--254 (2018)

\bibitem{muller2011smd}
Müller, M., Konz, V., Bogler, W., Arifi-Müller, V.: Saarland music data ({SMD}). Late-Breaking and Demo Session of the International Conference on Music Information Retrieval  (2011)

\bibitem{panda2020novel}
Panda, R., Malheiro, R., Paiva, R.P.: Novel audio features for music emotion recognition. IEEE Transactions on Affective Computing  \textbf{11}(4),  614--626 (2020)

\bibitem{panda2023feature}
Panda, R., Malheiro, R., Paiva, R.P.: Audio features for music emotion recognition: A survey. IEEE Transactions on Affective Computing  \textbf{14}(1),  68--88 (2023)

\bibitem{panda2013multi}
Panda, R., Malheiro, R., Rocha, B., Oliveira, A.P., Paiva, R.P.: Multi-modal music emotion recognition: A new dataset, methodology and comparative analysis. In: 10th International symposium on computer music multidisciplinary research (CMMR). pp. 570--582 (2013)

\bibitem{panda2011moodsvm}
Panda, R., Paiva, R.P.: Using support vector machines for automatic mood tracking in audio music. In: 130th AES Convention. vol.~1 (2011)

\bibitem{pasini2024musiclatent}
Pasini, M., Lattner, S., Fazekas, G.: Music2{L}atent: Consistency autoencoders for latent audio compression. In: Proc. 25th International Society for Music Information Retrieval Conference (ISMIR) (2024)

\bibitem{russell1980circumplex}
Russell, J.A.: A circumplex model of affect. Journal of personality and social psychology  \textbf{39}(6), ~1161 (1980)

\bibitem{Simonyan2014VeryDC}
Simonyan, K., Zisserman, A.: Very deep convolutional networks for large-scale image recognition. CoRR  \textbf{1409.1556} (2014)

\bibitem{Soleymani20131000SF}
Soleymani, M., Caro, M.N., Schmidt, E.M., Sha, C.Y., Yang, Y.H.: 1000 songs for emotional analysis of music. In: CrowdMM (2013)

\bibitem{song2012evalEmo}
Song, Y., Dixon, S., Pearce, M.: Evaluation of musical features for emotion classification. In: Proc. 13th International Society for Music Information Retrieval Conference (ISMIR) (2012)

\bibitem{tzanetakis2002marsyas}
Tzanetakis, G., Cook, P.: {MARSYAS}: a framework for audio analysis. Organised Sound  \textbf{4},  169--175 (2002)

\bibitem{won2024foundation}
Won, M., Hung, Y.N., Le, D.: A foundation model for music informatics. In: IEEE Int.~Conf.~on Acoustics, Speech and Signal Processing (ICASSP) (2024)

\bibitem{wu2014timbre}
Wu, B., Horner, A., Lee, C.: The correspondence of music emotion and timbre in sustained musical instrument sounds. AES: Journal of the Audio Engineering Society  \textbf{62},  663--675 (2014)

\bibitem{Yang_Chen_2011}
Yang, Y.H., Chen, H.H.: Music Emotion Recognition. CRC Press, Boca Raton, FL, USA (2011)

\bibitem{yang2008regression}
Yang, Y.H., Lin, Y.C., Su, Y.F., Chen, H.H.: A regression approach to music emotion recognition. IEEE Transactions on Audio, Speech, and Language Processing  \textbf{16}(2),  448--457 (2008)

\bibitem{Yang2006MusicEC}
Yang, Y.H., Liu, C.C., Chen, H.H.: Music emotion classification: a fuzzy approach. ACM Multimedia  (2006)

\bibitem{Zhang2018ThePD}
Zhang, K.J., Zhang, H., Li, S., Yang, C.Y., Sun, L.: The {PME}mo dataset for music emotion recognition. In: Proc. ACM on International Conference on Multimedia Retrieval (2018)

\end{thebibliography}
%





\end{document}